\documentclass[longauth,letter]{aa}  
\usepackage{graphicx}
\usepackage{stfloats}

\usepackage[varg]{txfonts}
\usepackage[breaklinks,pdfnewwindow=true,colorlinks=true,citecolor=blue]{hyperref}
\usepackage[version=3]{mhchem}
\usepackage{xspace}
\usepackage{orcidlink}
\newcommand{\orcid}[1]{\unskip\protect\href{https://orcid.org/#1}{\protect\includegraphics[width=8pt,clip]{logo_orcid}}}
\newcommand{\mjybm}{\ensuremath{\rm mJy\,beam^{-1}}\xspace}

\newcommand{\cc}{\ensuremath{\rm cm^{-3}}\xspace}
\newcommand{\kms}{{\ensuremath{{\rm km\, s^{-1}}}}\xspace}

\begin{document} 
   \title{Probing the Physics of Star-Formation (ProPStar) \thanks{Based on observations carried out under 
   project number S21AD with the IRAM NOEMA Interferometer and 
   090-21 with the IRAM 30m telescope. 
   IRAM is supported by INSU/CNRS (France), MPG (Germany) and IGN (Spain)}
   \thanks{The code to reproduce the results of the paper are hosted at 
   \url{github.com/jpinedaf/NGC1333_NOEMA_turbulence}}}

  \subtitle{III. No evidence for dissipation of turbulence down to 20 mpc (4\,000 au) scale}

   \author{
        Jaime E. Pineda\inst{1}\orcidlink{0000-0002-3972-1978}
        \and 
        Juan D. Soler\inst{2}\orcidlink{0000-0002-0294-4465}
        \and
        Stella Offner\inst{3}\orcidlink{0000-0003-1252-9916}
	\and
	Eric W. Koch\inst{4}\orcidlink{0000-0001-9605-780X}
        \and 
        Dominique M. Segura-Cox\inst{3,1,}\thanks{NSF Astronomy and Astrophysics Postdoctoral Fellow}\orcidlink{0000-0003-3172-6763}
        \and
        Roberto Neri\inst{5}\orcidlink{0000-0002-7176-4046}
        \and
        Michael Kuffmeier\inst{6,1}\orcidlink{0000-0002-6338-3577}
        \and
        Alexei V. Ivlev\inst{1}\orcidlink{0000-0002-1590-1018}
        \and
        Maria Teresa Valdivia-Mena\inst{1}\orcidlink{0000-0002-0347-3837}
        \and
        Olli Sipil\"a\inst{1}\orcidlink{0000-0002-9148-1625}
        \and
        Maria Jose Maureira\inst{1}\orcidlink{0000-0002-7026-8163}
        \and
        Paola Caselli\inst{1}\orcidlink{0000-0003-1481-7911}
        \and
        Nichol Cunningham\inst{7,8}\orcidlink{0000-0003-3152-8564}
        \and
        Anika Schmiedeke\inst{9}\orcidlink{0000-0002-1730-8832}
         \and
        Caroline Gieser\inst{1}\orcidlink{0000-0002-8120-1765}
	\and
        Michael Chen\inst{10}\orcidlink{0000-0003-4242-973X}
        \and
        Silvia Spezzano\inst{1}\orcidlink{0000-0002-6787-5245}
        }
        
\institute{
    Max-Planck-Institut f\"ur extraterrestrische Physik, Giessenbachstrasse 1, D-85748 Garching, Germany\\
    \email{jpineda@mpe.mpg.de}
    \and
    Istituto di Astrofisica e Planetologia Spaziali (IAPS), INAF, Via Fosso del Cavaliere 100, 00133, Roma, Italy
    \and
    Department of Astronomy, The University of Texas at Austin, 2500 Speedway, Austin, TX 78712, USA
    \and
    Center for Astrophysics | Harvard \& Smithsonian, 60 Garden St, Cambridge, MA 02138, USA
    \and
    Institut de Radioastronomie Millim\'etrique (IRAM), 300 rue de la Piscine, F-38406, Saint-Martin d'H\`eres, France
    \and
    Department of Astronomy, University of Virginia, Charlottesville, VA, 22904, USA
    \and
    SKA Observatory, Jodrell Bank, Lower Withington, Macclesfield SK11 9FT, United Kingdom
    \and
    IPAG, Universit\'{e} Grenoble Alpes, CNRS, F-38000 Grenoble, France
    \and
    Green Bank Observatory, PO Box 2, Green Bank, WV 24944, USA
    \and
    Department for Physics, Engineering Physics and Astrophysics, Queen's University, Kingston, ON, K7L 3N6, Canada
    }

   \date{Received Month Day, Year; accepted Month Day, Year}

 
  \abstract
   {Turbulence is a key component of molecular cloud structure.
   It is usually described by a cascade of energy
   down to the dissipation scale. 
   The power spectrum for subsonic incompressible turbulence is $\propto k^{-5/3}$, 
   while for supersonic turbulence it is $\propto k^{-2}$.}
   {We aim to determine the power spectrum in an actively star-forming 
   molecular cloud, from parsec scales down to the expected magneto-hydrodynamic 
   (MHD) wave cutoff (dissipation scale).}
   {%
   We analyze observations of the nearby NGC 1333 star-forming region 
   in three different tracers to cover the different scales from $\sim$10 pc down to 20 mpc. 
   The largest scales are covered with the low density gas tracer 
   \ce{^{13}CO} (1--0) obtained with single dish, 
   the intermediate scales are covered with single-dish observations 
   of the \ce{C^{18}O} (3--2) line, while the smallest scales 
   are covered in \ce{H^{13}CO+} (1--0) and \ce{HNC} (1--0) with a combination of 
   NOEMA interferometer and IRAM 30m single dish 
   observations. 
   The complementarity of these observations enables us to 
   generate a combined power spectrum covering more than two orders of 
   magnitude in spatial scale. 
   }
   {We derive the power spectrum in an active 
   star-forming region spanning more than 2 decades 
   of spatial scales. 
   The power spectrum of the intensity maps shows a 
   single power-law behavior, with an exponent of 
   $2.9\pm 0.1$ and no evidence of dissipation. 
   Moreover, there is evidence for the power-spectrum of the ions 
   to have more power at smaller scales than the neutrals, which 
   is opposite from theoretical expectations.
   }
   {We show new possibilities of studying the 
   dissipation of energy at small scales in star-forming 
   regions provided by interferometric observations.
   }

   \keywords{astrochemistry; 
             Turbulence;
             ISM: molecules; 
             ISM: clouds; 
             stars: formation;
             ISM: individual objects: NGC 1333
             }

   \maketitle
%

\section{Introduction}%
Dense cores are the places where stars form \citep{Pineda2023-PP7}. 
These are the places with subsonic levels of turbulence \citep{Goodman1998-Transition_to_Coherence,Pineda2010-B5_Transition_to_Coherence,Friesen2017-GAS_DR1}, 
and this transition to coherence has been discussed as possibly being linked to the dissipation of turbulence.

Turbulence in clouds has been typically studied using different 
tracers, which probe different scales of the cloud.
For example, \cite{Larson1981} found a correlation between 
linewidth and cloud size, $\sigma_v \propto L^{0.38}$. 
Further analysis using principal component analysis (PCA) showed 
that the velocity structure function as a function of spatial scale, $l$, 
is universal in molecular clouds  \citep{HeyerBrunt04}, 
$\delta v \propto l^{0.49}$, 
although further work showed some small 
variation \citep{Heyer2009-Revisit_Larson_Law_GMC,Roman-Duval2011-Turbulence_MCs_GRS}
the relation stands more or less universal 
\citep[see also references in ][]{McKeeOstriker07,HeyerDame15}.

In particular, the spatial power spectrum is used 
to study the lower density section of the cold neutral medium.
\cite{Miville-Deschenes2010-Herschel_Polaris_turbulence} measured the spatial power spectrum in 
Polaris cloud using {\it Herschel} data, 
while \cite{Miville-Deschenes2016-Turbulence_Cirrus} showed 
a lack of evidence for the cutoff. 
These works focused on cirrus clouds, where gravity is not expected 
to play a major role.

In the case of star-forming molecular clouds, 
various works predict a turnover in the power spectrum at the dissipation scale 
\citep[e.g.,][]{ElmegreenScalo04}.
Moreover, in the case of magnetohydrodynamic (MHD) turbulence,
ions and neutrals at higher angular resolution are expected to display a difference in their linewidths 
at the same characteristic scale $l$ \citep{LiHoude2008-Turbulence_Dissipation_M17}. 
This is because there is a minimum MHD wave 
for wave propagation ($\lambda_{min}$), below which MHD waves will not transmit, 
which would imply that ions should exhibit a turnover in the power-spectrum at larger scales than neutrals.

In this study, we present a comprehensive analysis of the power spectrum 
of spatial frequencies of gas in the 
NGC 1333 young cluster in the Perseus molecular cloud. 
We combine three data sets to study the power spectrum from 
parsec scales down to 20 mpc (4\,000 au).

\section{Data}
In order to properly probe the power spectrum over a wide 
spatial dynamical range we use three different data sets with different tracers.
This enables us to determine the power spectrum 
over an extended range of physical scales, which combined enable us to cover more than 
2 decades of physical scales. 
Figure~\ref{fig:data} shows  the integrated intensity maps used to analyze the scales covered, 
as well as the footprint of the higher angular resolution (and density tracer) observations used for the 
next spatial scale.
The median spectra for the different data sets are presented in Sec.~\ref{app:median_spectra} and 
shown in Fig.~\ref{fig:median_spectra}.

\subsection{Molecular Cloud Tracers}
For the largest scale, we use the \ce{^{13}CO} (1--0) map of Perseus obtained  
with the FCRAO telescope by the COMPLETE survey
\citep{Ridge2006-COMPLETE}. 
This data has a resolution of 46$\arcsec$ and a noise level of 0.12 K.

\begin{figure*}
    \centering
    \includegraphics[width=\textwidth]{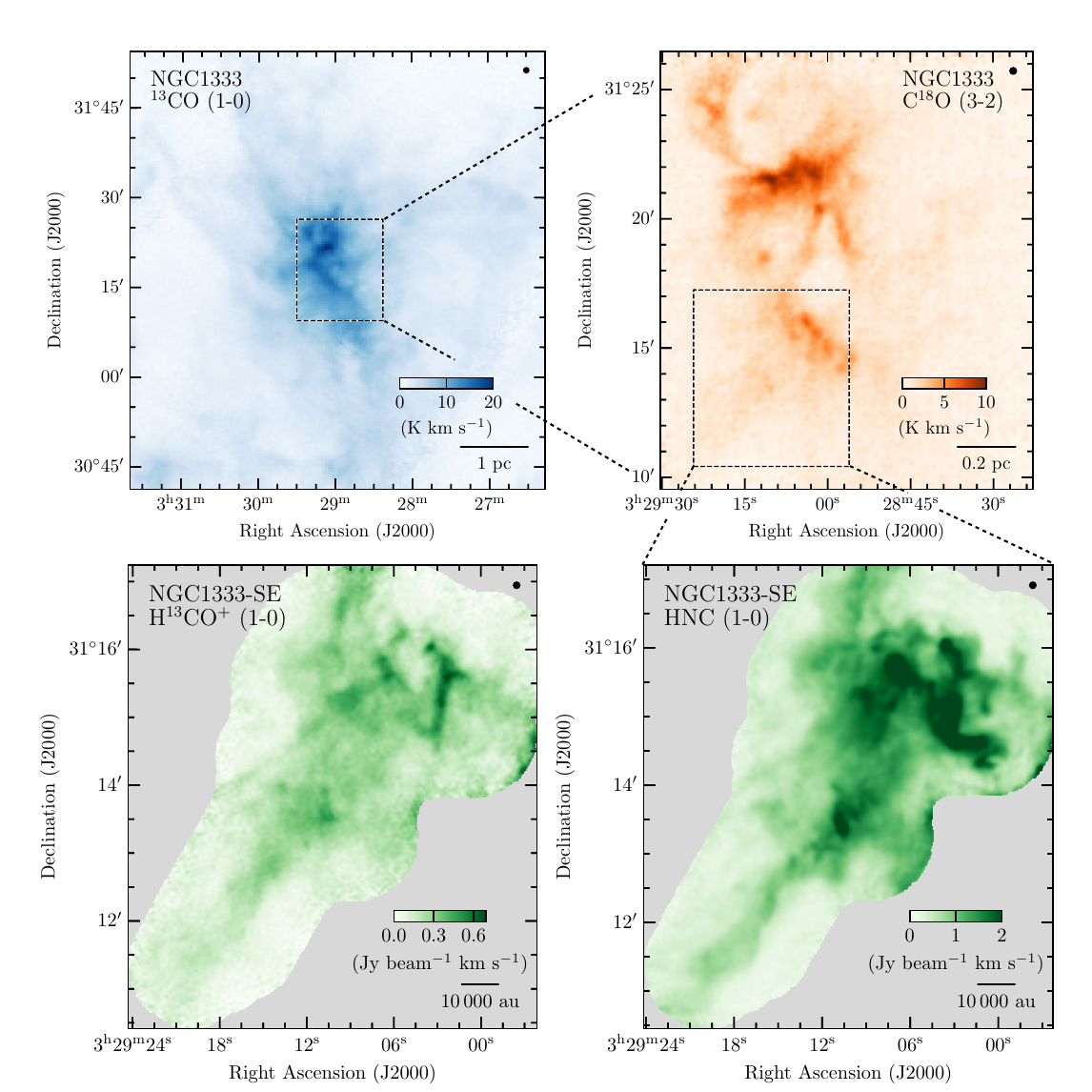}
    \caption{Integrated intensity map of the different tracers used to observe the NGC1333 region.
    {\it Top left:} Large scale is traced with \ce{^{13}CO} (1--0). 
    The footprint of the \ce{C^{18}O} (3--2) map is marked by
    the dashed box. 
    {\it Top right:} Intermediate scale is traced with 
    \ce{C^{18}O} (3--2). 
    The footprint of the interferometric data is marked by
    the dashed box. 
    {\it Bottom left and right:} Highest angular resolution maps 
    of \ce{H^{13}CO+} and \ce{HNC} (1--0), respectively. 
    The beam size and scale bar are shown in the top right and bottom right corners, respectively.
    }
    \label{fig:data}
\end{figure*}

The integrated intensity is calculated by adding the emission in the velocity range of $-1$ and $11.5$ \kms.
The top left panel of Fig.~\ref{fig:data} shows the 
integrated intensity map.

For intermediate scales, we use
the \ce{C^{18}O} (3--2) observations of NGC 1333 
taken with HARPS at JCMT \citep{Curtis2010-JCMT_Perseus_CO}.
The angular resolution of these observations is 17.7$\arcsec$  
and it has a noise level of 0.18 K \citep[][]{Curtis2010-JCMT_Perseus_CO}.

The integrated intensity
maps are calculated between 5.5 and 10 \kms, which covers all the
emission seen. 
The top right panel of Fig.~\ref{fig:data} shows the 
integrated intensity map.

To trace small scales, we select \ce{HNC} and \ce{H^{13}CO+} (1--0), because they are the brightest pair of molecular lines 
tracing the cloud, and have little contamination from outflows or shocks.
They both cover a large fraction of the map and 
do not display resolved hyperfine components. 
The details on the image processing for these data are presented in Section~\ref{sec:IRAM_data}.

\section{Analysis}
\subsection{Intensity Power Spectrum}
The power spectrum of the integrated intensities is calculated using 
the {\em Spatial Power Spectrum} function, $PS_2$, in 
\verb+Turbstat+ \citep{Koch2019-TurbuStat} on the 
integrated intensity maps of the different transitions used, 
see Table~\ref{tab:spectral_setup}.
We perform a beam correction on the power spectrum available on \verb+Turbstat+, 
and in addition we only use spatial frequencies down to $3\times$ the beam size, 
which allows us to avoid the regime where the beam affects the power spectrum measurements. 
We also apodize the images with the \verb+CosineBell+ kernel,
which is available in \verb+Turbstat+, 
to reduce the ringing effects due to the images edge\footnote{\url{https://turbustat.readthedocs.io/en/latest/tutorials/applying_apodizing_functions.html}}.

In the case of the interferometric observations, we pad the region 
not covered by the mosaic to reduce the effects from the map coverage on the 
power spectrum. 
The maps are filled with Gaussian noise outside the mosaic with a sigma level of the value listed in Table~\ref{tab:spectral_setup}.

In this analysis we use spatial frequency, $k=1/\lambda$, 
where $\lambda$ is the spatial scale probed.

\subsection{Stitching Different Scales and Power-law Fit}
\begin{figure*}
    \centering
    \includegraphics[width=0.49\textwidth]{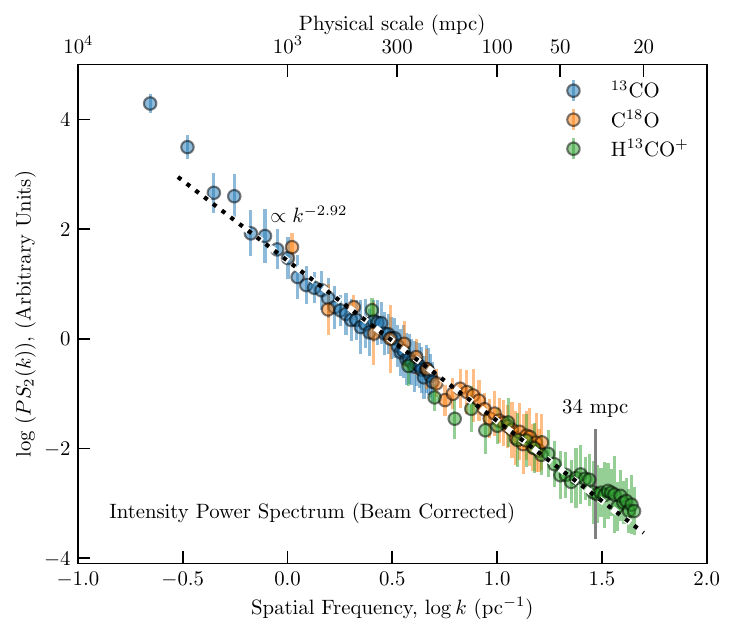}
    \includegraphics[width=0.49\textwidth]{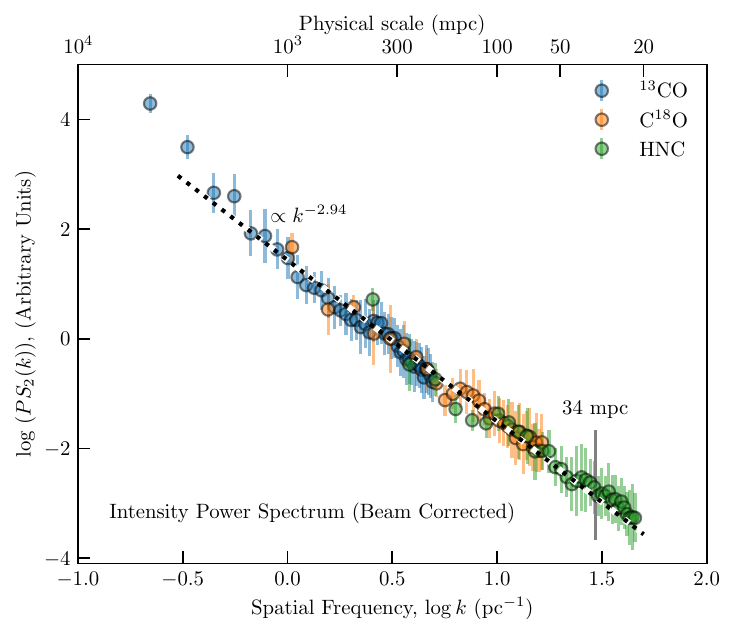}
    \caption{Combined power spectrum of the three different datasets.
    Left and Right panels show the power spectrum with ion and neutrals at the smallest scales, respectively.
    The estimated MHD cutoff scale, 34 mpc, is marked by a vertical line.
    The fitted power law is shown by the dotted line, with exponents of {$\beta=2.92\pm0.11$} and {$2.94\pm0.11$} 
    for the left and right panels, respectively.}
    \label{fig:TdV_PowerSpetrum}
\end{figure*}

The individual $PS_2$ from each molecular line have a different absolute amplitude, 
due to the difference in relative abundances and volumes traced by each molecular transition.
However, the power-law exponent is inter-comparable.
Therefore, we apply a normalisation coefficient to the $PS_2$ derived for each map.
We normalize the \ce{C^{18}O} power spectrum, $PS_{2,\, \ce{C^{18}O}}(k)$, to 1 at a spatial 
frequency of $10^{-0.5}$ pc$^{-1}$, while the normalization parameter, 
$f_{mol}$, 
is added to the spatial power spectrum, $\log{PS_{2, \textrm{mol}}} + f_{\textrm{mol}}$, 
of each molecular transition as a free parameter, where $mol$ is 
\ce{^{13}CO}, \ce{H^{13}CO+}, and \ce{HNC}, respectively. 
In practice, we generate two different dataset for analysis: 
$D_1 = \{$\ce{^{13}CO}, \ce{C^18O}, and  \ce{H^{13}CO+}$\}$ and 
$D_2 = \{$\ce{^{13}CO}, \ce{C^18O}, and  \ce{HNC}$\}$.
{We fit a single power-law to each of the combined datasets as
\begin{equation}
\log PS_{2, i}(k) =  
A_{i} - \beta_{i} \log k~,
\end{equation}
where $k$ is the spatial frequency in units of pc$^{-1}$, %
$\beta_i$ is the power-spectrum power-law index and $A_i$ sets the amplitude of the relation 
fitted to the dataset $D_i$.
The power-laws and all the normalization parameters are fitted simultaneously 
using \texttt{EMCEE} \citep{emcee}, 
see Appendix~\ref{sec:appendix_fit} for details.
}
The results are shown in Fig.~\ref{fig:TdV_PowerSpetrum} while the fitted parameters are listed in Table~\ref{tab:power-law_fit}. 

\begin{table}
\caption{Results of the power-law fits.\label{tab:power-law_fit}}
\renewcommand{\arraystretch}{1.5} %
{
\begin{center}
\begin{tabular}{lc}
\hline \hline
Parameter & Value \\
\hline
{$A_{\ce{H^{13}CO^+}}$} & $1.42_{-0.10}^{+0.10}$ \\
{$\beta_{\ce{H^{13}CO^+}}$} & $2.92_{-0.11}^{+0.11}$ \\
{$A_{\ce{HNC}}$} & $1.43_{-0.10}^{+0.10}$ \\
{$\beta_{\ce{HNC}}$} & $2.94_{-0.11}^{+0.11}$ \\
{$f_{\rm ^{13}CO}$} & $2.51_{-0.08}^{+0.08}$ \\
{$f_{\rm H^{13}CO^+}$} & $-0.25_{-0.11}^{+0.11}$ \\
{$f_{\rm HNC}$} & $-1.50_{-0.11}^{+0.10}$ \\
\hline
\end{tabular}
\end{center}}
\end{table}

\section{Discussion}

The physical interpretation of the intensity power spectrum relies on 
the assumption that the emission is optically thin, and 
therefore it is the power spectrum of the column density. 
Then the power spectrum is divided into three different 
regimes \citep[e.g.,][]{Federrath+10}.
At large scales, the transition between a plateau and a power law is associated with the injection scale of turbulence 
(usually expected to be similar to the clouds physical scale, \citealt{McKeeOstriker07}).
At intermediate scales, the power law is associated with the inertial range, that is, the range of scales that characterize self-similarity in the flow. 
However, features at smaller scales could appear due to the influence of 
outflows or other feedback mechanisms \citep{NakamuraLi05,Padoan2009-PowerSpectrum_NGC1333,Boyden2016-Feedback_Turbulence,Xu2020-Simulation_Bubbles_CNN}.
At small scales, the change in the slope of the power law is associated with the dissipation scale in the turbulence cascade.
Our data appears to be beyond the range of the injection scales, 
but in what follows, we discuss the other two features of the power spectra.

\subsection{Exponent of the Power Spectra, $\beta$}

We obtain a power spectrum with an average exponent of {$\beta=2.9\pm 0.1$},
with the values of {$2.92\pm0.11$} and {$2.94\pm0.11$} 
for \ce{H^{13}CO+} and \ce{HNC}, respectively.
This value is in agreement with the exponent found by 
\cite{Padoan2009-PowerSpectrum_NGC1333} on the same region using \ce{^{13}CO} (1--0) 
\citep[see also][]{Padoan2006-PowerSpectrum_Perseus_Simple}, 
and it is also in rough agreement with the exponents found 
in \ion{H}{i}, \ce{^{12}CO} (1--0), \ce{^{13}CO} (1--0), and 
dust extinction toward the Perseus molecular cloud \citep{Pingel2018-PowerSpectrum_Perseus_Multitracer}.
However, \cite{Pingel2018-PowerSpectrum_Perseus_Multitracer} 
matches the pixel to the beam size and does not apply a beam 
correction in the power spectrum.
This does not have an effect at large scales, however, it is 
important at scales closer to the beam and 
complicates a detailed comparison with the indexes obtained 
by \cite{Pingel2018-PowerSpectrum_Perseus_Multitracer}.
In other molecular clouds, studies with single-dish observations 
also recover an exponent comparable to the one determined here
\citep{Stutzki1998-Fractal_MCs,Sun2006-Perseus_KOSMA,Xu2020-Simulation_Bubbles_CNN}.
Similar efforts focused on the power spectrum 
of the intensity taking advantage of interferometric mosaics combined 
with single-dish observations to recover all scales.
In L1551, \cite{Swift2008-L1551_Mosaic_CaseStudy}  found an exponent 
of 2.8$\pm$0.1 using \ce{C^{18}O} (1--0);  
while in different subregions of Orion~A, 
\cite{Feddersen2019-CARMA_NRO_Orion_Turbulence} 
found exponents between 3 and 4 using \ce{^{12}CO} and \ce{^{13}CO} (1--0), 
while slopes between 2.2 and 3.4 are determined in the case of \ce{C^{18}O} (1--0). 
In these cases, the most optically thin transition used is 
\ce{C^{18}O} (1--0), which still should become optically thick 
close to dense cores or embedded YSOs, and gives exponent values 
comparable to those found in this work. 
{We note that the wide range of exponents reported in the literature complicates our comparison. 
This variation could suggest that the power-law exponent depends on the local environment, 
or it may indicate that large uncertainties in previous studies, possibly due to optical depth issues, 
are affecting the results.}

Our derived slopes match those derived
by \cite{Miville-Deschenes2016-Turbulence_Cirrus} from their combined 
spectrum ranging from scales between 0.01 and 50\,pc ($2.9\pm0.1$).
However, the region considered in \cite{Miville-Deschenes2016-Turbulence_Cirrus} is a Galactic cirrus, 
with densities orders of magnitude lower 
than those characteristic of the gas in NGC 1333.
Their $2.9 \pm 0.1$ exponent is comparable to that found in the infrared emission from cirrus observed with the Infrared Astronomical Satellite (IRAS) 
at 100\,$\mu$m and {\it Herschel} at 250, 350, and 500\,$\mu$m \citep{Gautier1992-Noise_Cirrus,Miville-Deschenes2010-Herschel_Polaris_turbulence}.
However, other Galactic cirrus clouds present different exponents 
in the power spectra derived from neutral atomic hydrogen (\ion{H}{i}),
with exponents between $2.59$ and $3.7$ on the largest scales 
\citep[e.g.,][]{Pingel2013-Turbulence_Translucent_Cloud,Martin2015-GHIGLS_Dana_and_Power_Spectrum}.
Similarly, the power spectra of the dust thermal emission observed 
with {\it Herschel} toward the Large Magellanic cloud show 
exponents between $1.0$ and $2.43$ \citep{Colman2022-Turbulence_Driving_ISM}.
This variability is found in multiple studies across tracers \citep{Szotkowski2019-HI_PowerSpectrum_LMC,Koch2020-PowerSpectrum_LocalGroup}, implying that the power law exponent 
at large scales is not universal 
but is related to local physical conditions.

Numerical simulations of hydrodynamic (HD) and 
magneto-hydrodynamic (MHD) turbulence are commonly used to 
link the power spectrum exponent to physical quantities, 
such as the sonic Mach number ($\mathcal{M}_{s}$), 
the Alfv\'{e}n Mach number ($\mathcal{M}_A$), 
and the mixture between solenoidal, and compressive turbulent modes 
\citep[e.g.,][]{FederrathKlessen2013-SFE_Turbulent_Cloud,Burkhart2015-Turbulence_GMCs_Diagnostics}.
\cite{Kim2005-PowerSpectrum_HD_Exponent} used numerical simulations of turbulence in a compressible and isothermal medium to show that trans-sonic turbulence 
($\mathcal{M}_{s}\approx 1$) produces power spectra with exponents around 1.73, close to the expected values from Kolmogorov turbulence, 
and shallower for increasing $\mathcal{M}_{s}$.
A power spectrum slope of $\approx3.0$  corresponds to 
their results for $\mathcal{M}_{s}\approx 7$.

Dedicated studies of MHD turbulence in self-gravitating 
giant molecular clouds show power spectrum exponents between 
$1.5$ and $2.5$ for simulations with $0.5 < \mathcal{M}_{s} < 20$ 
and $0.7 < \mathcal{M}_{A} < 2.0$ 
\citep{Burkhart2015-Turbulence_GMCs_Diagnostics}.
These slopes are hard to reconcile with our observational results and strongly suggest the effect of additional physics.
\cite{Boyden2016-Feedback_Turbulence} include the effect of stellar winds and magnetic fields in the MHD simulations aimed at reproducing the physical conditions in the Perseus molecular cloud and produce synthetic \ce{^{12}CO} (1--0) emission observations. 
However, the slopes of their resulting power spectra are lower than in our results, falling in the range between $2.64$ and $2.77$. 
\cite{Boyden2018} showed that larger variations in temperature and abundance tend to flatten the power spectrum slope, 
and synthetic observations of MHD turbulence produced using full astrochemical networks yield slopes of $\sim$2.8.

Optical depth has a strong effect on the derived slope, such that the power spectrum slope of 
optically thick turbulent gas limits to $3.0$ \citep{LazarianPogosyan2004-PowerSpectrum,Boyden2018}.
However, this is not the case in our observations, 
where the \ce{C^{18}O} and the higher-density tracers 
do not show evidence of being optically thick.
Given the large degeneracy between the physical parameters that can result in our observed slope, we focus on other aspects of our observations, namely the similarity between the results for neutral and ionized species and the absence of a cut-off frequency in the power spectra.

{Fig.~\ref{fig:TdV_PowerSpetrum} and \ref{fig:Ion-Neutral} show that there is a feature at
$\log k \approx$0.9 pc$^{-1}$ ($\approx$150 mpc) there is a feature in the power spectra.  
However, this feature is not in the \ce{C^{18}O}  power-spectrum, which covers 
the same scales and with more independent samples. 
Therefore this feature could be related to the shape of the interferometric mosaic, since the 
90$\arcsec$ scale (corresponding to $\log k \approx$0.9 pc$^{-1}$) is comparable to the 
width of the mosaic covering the south-east filamentary structure.
}

\subsection{Lack of difference between Ions and Neutrals}

The neutral and ionized species are expected to decouple at certain scales due to 
ambipolar diffusion \citep[][]{MestelSpitzer56,Kulsrud1969-Ambipolar_Diffusion_CosmicRays,Mouschovias2011-MHD}. 
This decoupling should introduce differences in the emission power spectra between neutral and ionized tracers 
\citep[e.g.,][]{Houde2000-Bfield_Ion_Spectra}.

\begin{figure}[h]
    \centering
    \includegraphics[width=\columnwidth]{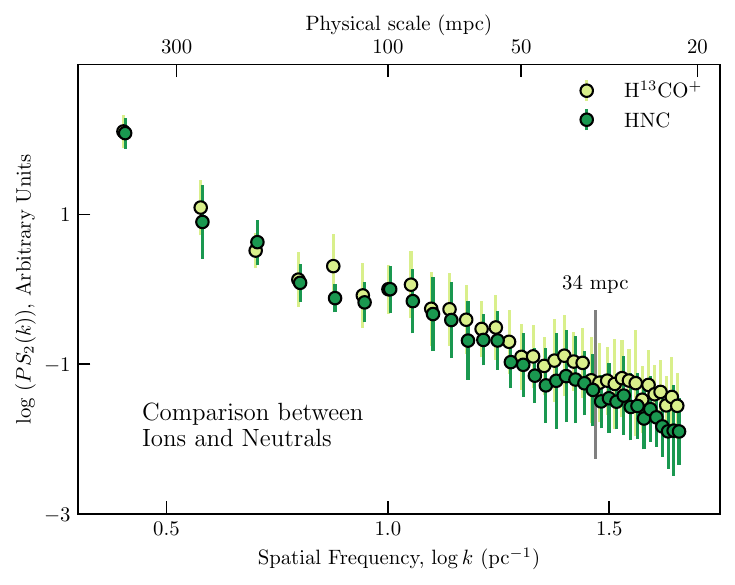}
    \caption{Comparison of power spectra of ions and neutrals 
    with our highest angular resolution data. 
    The spectra are normalized as {$PS_2(k)=1$} at the spatial frequency 
    of 10 pc$^{-1}$.
    The ions have a subtle but systematic excess of power at 
    the smallest scales when compared to the neutrals.
    This is opposite to the expectations from ambipolar diffusion.
    }
    \label{fig:Ion-Neutral}
\end{figure}

The comparison of the complete power spectra between ions and neutrals 
(see Fig.~\ref{fig:TdV_PowerSpetrum}) shows that they present the same 
power-law and  neither has a turnover.
This suggests that ions and neutrals do not present a substantial difference down to 20 mpc.
Although both spectra agree well within 1-$\sigma$, there is a systematic difference 
between the power spectra.
Fig.~\ref{fig:Ion-Neutral} shows this {possible} offset appears between the power spectra 
of {\ce{HNC}} and \ce{H^{13}CO+} for the highest angular resolution data.
This is the opposite of the prediction from ambipolar diffusion theories 
\citep[e.g.,][]{LiHoude2008-Turbulence_Dissipation_M17}.
Similarly, we cannot rule out that this discrepancy in the power spectra is related to systematic differences in the 
abundance and/or excitation conditions at the higher angular scales
\citep{Gaches2015-Astrochemistry_Simulations_MC,Pineda2022-HMM1_NH3_Depletion,Tritsis2023-Ion_Neutral_Drift_Observe}.

\subsection{MHD wave cutoff}
In a molecular cloud, where the ionization level is low, Alfv\'en waves cannot propagate 
when the collision frequency between ions and neutrals is comparable 
to or smaller than the MHD wave frequency \citep{Kulsrud1969-Ambipolar_Diffusion_CosmicRays,Mouschovias2011-MHD,Hennebelle2023-Dust_Inertia_Waves}. 
Therefore, there is a critical length for wave propagation beyond which waves do not transmit.
The critical length for wave propagation 
\citep{StahlerPalla-2005,Mouschovias2011-MHD} 
is obtained from,
\begin{equation}
\lambda_{min} = 
\sqrt{\frac{\pi}{4 \mu \, m_{\ce{H}} \, n(\ce{H2})}} \frac{B_o}{n(\ce{H2})\, X(e)\, 
\langle \sigma\, v \rangle~}, 
\end{equation}
where $B_o$ is the unperturbed magnetic field, and 
$\langle \sigma\, v \rangle$ is the rate coefficient for elastic collisions.
The rate coefficient term is approximated by the Langevin term,
\begin{equation}
\langle \sigma\, v \rangle = 1.69\times 10^{-9}\, {\rm cm^3\,s^{-1}}~,
\end{equation}
for \ce{HCO+}--\ce{H2}  collisions 
\citep{McDaniel:1973vs}. 
We estimate the magnetic field strength, $B_o$, using the 
relation of \cite{Crutcher2010-B_field_Bayesian} \citep[see also][]{Myers2021-Bfield_Dense_Cores,Pattle2023-PP7_Bfield},
a typical volume density of $10^{3.6}$ \cc, 
and the median value of $X(\ce{e})=10^{-6.5}$ reported by \cite{Pineda2024-ProPStar_NGC1333_CRIR}.
With these values, we obtain an MHD wave cutoff scale 
of 0.034 pc (7\,000 au). 
Therefore, we predict that if the turbulence is 
MHD in nature, then a dissipation scale should be observed at 
$\approx$0.034 pc (or $\approx$37$\arcsec$ at the distance of Perseus).

Previously, this scale has been proposed as a 
possible origin for the transition to coherence 
between the supersonic cloud and subsonic cores 
\citep[$\sim$0.1 pc;][]{Goodman1998-Transition_to_Coherence,Caselli2002-Cores_N2Hp,Pineda2010-B5_Transition_to_Coherence}. 
Recent observations of dense gas with different 
tracers (e.g., \ce{N2H+} and \ce{NH3}) and 
angular resolutions, showed that the filaments present in the 
southern end of NGC 1333  presents subsonic levels of 
turbulence at scales of $\approx 40\arcsec$
\citep{Friesen2017-GAS_DR1,Hacar2017-NGC1333,Dhabal2019-NGC1333_VLA,Sokolov2020-NGC1333_Bayesian},
which corresponds to 0.034 pc or 7\,000 au at the distance 
of Perseus.

Finally, an improved estimate of the density or magnetic fields strength are needed to improve the obtained constrains. 
We use the volume density map of the region \citep{Pineda2024-ProPStar_NGC1333_CRIR} to determine a 
mean density $10^{3.6\pm0.2}$ \cc, and with this uncertainty we derive an uncertainty on the 
MHD wave cutoff of $\pm 13$ mpc. 
This would push the $\lambda_{min}$ right below the scales sampled by these observations. 
Therefore, higher angular resolution and tighter constrains on the density and magnetic field strength 
in the region are required to make more progress in this topic.

\subsection{Interpretation of single power-law spectrum}
The combination of previous evidence for ions presenting 
a higher level of turbulence than neutrals at small scales 
suggests that a more exotic physical process is at play.

\cite{Pineda2021-B5_Ions_Neutral} suggested that MHD waves could 
penetrate within dense cores and perturb the magnetic field lines. 
As a result, this process would inject kinetic energy 
into the ions at smaller scales, and therefore it could remove the scale for 
dissipation and possible differences between ions and neutrals.

Another possibility proposed by \cite{Hennebelle2023-Dust_Inertia_Waves}
involves the effect of the dust grain inertia on the transfer of energy at 
smaller scales. 
This mechanism involves the interaction between dust particles and gas 
at small scales, which could inject energy in the denser regions remove 
the scale for dissipation.

On the other hand, different  two-fluid simulations have studied the nature of the turbulence of both ions and neutrals 
\citep{Oishi2006-Turbulence_AD_smallscale,Burkhart2015-Alfvenic_Turbulence_AD_Scale,Hu2023-Ion_Neutral_MHD_Damping},
however, these results show that the energy can be transported across the ambipolar diffusion (AD) scale. 
These results would suggest that our understanding of the AD process in more realistic conditions is 
incomplete.

Unfortunately, all these possible explanations have not provided 
synthetic observations to better compare with the different
observations, leaving the door open for improved comparisons. 
This includes possible effects due to radiative transfer and/or 
chemistry that must be taken into account to make the comparison.

\section{Conclusions}
We study the molecular cloud structure 
across different densities and scales by combining three different data sets. 
We derive the combined power spectrum %
covering more than 2 orders of magnitude (from $\approx$3 pc to 20 mpc) 
in linear scale, 
with the smallest scales probed by two tracers: 
\ce{H^{13}CO+} and \ce{HNC}.
The combined power spectrum {fitting across  all scales} is well fit with a 
{$PS_2(k) \propto k^{-\beta}$} function, where the exponents 
are {$2.92\pm0.11$} and {$2.94\pm0.11$} for 
\ce{H^{13}CO+} and \ce{HNC}, respectively.
The power spectrum shows no evidence of 
a turnover (dissipation)  down to 20 mpc. 
    
The power spectra of the ions and neutrals are not substantially different. 
However, a systematic offset (extra) power is present in the ions at 
small scales when compared to the neutrals, which 
is opposite to the expectations of ambipolar diffusion models.
These suggestive results, combined with the previous observations 
showing broader line-widths for ions compared to neutrals \citep{Pineda2021-B5_Ions_Neutral},
imply that contrary to the classical picture of 
turbulence dissipation our understanding of AD processes is incomplete.
More theoretical and observational work is needed 
\citep[e.g.,][]{Hennebelle2023-Dust_Inertia_Waves,Hu2023-Ion_Neutral_MHD_Damping}
to explain these results and provide a general framework for ion/neutral turbulence {and to 
further confirm these differences}.

\begin{acknowledgements}
    Part of this work was supported by the Max-Planck Society. 
    {We kindly thank the anonymous referee for the comments that helped improve the manuscript.} 
    EWK acknowledges support from the Smithsonian Institution as a Submillimeter Array (SMA) Fellow and the Natural Sciences and Engineering Research Council of Canada.
    DMS is supported by an NSF Astronomy and Astrophysics Postdoctoral Fellowship under award AST-2102405.
    JEP thanks {M. A. Miville-Desch\^enes and F. Boulanger for stimulating discussion when starting the project, and} 
    P. Hennebelle, A. Traficante, and S. Molinari 
    for valuable discussions.
    The authors acknowledge Interstellar Institute's program With Two Eyes and the Core2disk-III residential program of Institut Pascal at Universit\'e Paris-Saclay, with the support of the program ``Investissements d'Avenir'' ANR-11-IDEX-0003-01.
    This work made use of Astropy:\footnote{http://www.astropy.org} a community-developed core Python package and an ecosystem of tools and resources for astronomy 
    \citep{astropy:2013, astropy:2018, Astropy2022-v5}.
    This research has made use of NASA's Astrophysics Data System Bibliographic Services.
\end{acknowledgements}

\bibliographystyle{aa}
\bibliography{NGC1333_NOEMA_turbulence}
\label{LastPage} 

\begin{appendix}

\section{Processing of IRAM Data\label{sec:IRAM_data}}

The \ce{HNC} (1--0) and \ce{H^{13}CO+} (1--0) lines  were observed with IRAM 30-m and NOrthern Extended Millimeter Array (NOEMA) and 
the details are described  in Sec.~\ref{subsec:30m} and \ref{subsec:NOEMA}, respectively. 
The combination procedure is described in Sec.~\ref{subsec:combine}.

\begin{table*}[b!]
\caption{Spectral information for each of the spectral lines analyzed.\label{tab:spectral_setup}}
\begin{tabular}{lccccccc}
\hline \hline
Transition & Rest Freq. & Beam (PA) & rms & rms unit & $V_{min}$ & $V_{max}$ & Padding noise\\
 & (MHz) & (\arcsec$\times$\arcsec) & & & (\kms) & (\kms) & (\mjybm\,\kms)\\
 \hline
\ce{^{13}CO} (1--0) & 110201.35430 & 46 & 0.12 & K & $-$1.0 & 11.5 & --\\
\ce{C^{18}O} (3--2) & 329330.55250 & 17.7 & 0.18 & K & 5.5 & 10 & --\\
\ce{HNC} (1--0) & 90663.568 & 4.9$\times$4.7($-$38$\degr$) & 23 & \mjybm & 3.5  & 11 & 30 \\
\ce{H^{13}CO+} (1--0) & 86754.2884 & 5.0$\times$4.9($-$43$\degr$)  & 15 & \mjybm  & 5.3  & 10 & 20 \\
\hline
\end{tabular}
\end{table*}

\subsection{IRAM 30m telescope\label{subsec:30m}}
The observations were carried out with the IRAM 30m telescope at Pico Veleta (Spain) on 
2021 November 9, 10, and 11; and 
2022 February 19, 20, under project 091-21.
The EMIR E090 receiver and the FTS50 backend were employed. 
We used two spectral setups to 
cover the \ce{H^{13}CO+} (1--0) and \ce{HNC} (1--0)  
lines at 86.7 and 90.6 GHz (see Table \ref{tab:spectral_setup}).
We mapped a region of $\approx$150$\arcsec\times$150$\arcsec$ with the On-the-Fly mapping technique, and using position switching.
Data reduction was performed using the CLASS program of the GILDAS package\footnote{\url{http://www.iram.fr/IRAMFR/GILDAS}}. 
The beam efficiency, $B_{eff}$, is obtained using the Ruze formula 
(available in CLASS), and it is used to convert the observations 
into main beam temperatures, $T_{\rm mb}$.

\subsection{NOEMA interferometer\label{subsec:NOEMA}}
The observations were carried out with the IRAM NOEMA interferometer within the S21AD program 
using the Band 1 receiver.
The observations were carried out on 2021 
July 18, 19, and 21; 
August 10, 14, 15, 19, 22, and 29; and 
September 1
in the D configuration.
We observed a total of 96 pointings, which were separated 
on four different scheduling blocks.
The mosaic's center is located at 
$\alpha_{J2000}$=03$^{\rm h}$29$^{\rm m}$10.2$^{\rm s}$, 
$\delta_{J2000}$=31$\degr$13$\arcmin$49.4$\arcsec$.
We use the PolyFix correlator with a LO frequency of 82.505\,GHz and an instantaneous
bandwidth of 31 GHz spread over two sidebands (upper and lower) and two polarisations.
The centers of the two 7.744\,GHz wide sidebands are separated by 15.488\,GHz. 
Each sideband is composed of two adjacent basebands of $\sim$3.9\,GHz width (inner and outer basebands). 
In total, there are thus eight basebands, which are fed into the correlator. 
The spectral resolution is 2\,MHz throughout the 15.488\,GHz effective bandwidth per polarization. 
Additionally, a total of 112 high-resolution chunks are placed, 
each with a width of 64\,MHz and a fixed
spectral resolution of 62.5\,kHz. 
Both polarizations (H and V) are covered with the same spectral setup, and therefore the high-resolution chunks provide 66 dual polarisation spectral windows.
The high spectral resolution windows used here are listed in Table~\ref{tab:spectral_setup}.

\subsection{Image Combination\label{subsec:combine}}
The original IRAM 30m data is resampled to match the spectral setup 
of the NOEMA observations. 
We use the task \verb+uvshort+ to generate the pseudo-visibilities 
from the 30-m data for each NOEMA pointing. 
The imaging is done with natural weighting, a support mask, and 
using the SDI deconvolution algorithm \citep{Steer1984-SDI_Clean}.

The noise level of the combined images is reported in Table~\ref{tab:spectral_setup}. 
The integrated intensity
maps are calculated using the velocity ranges listed in Table~\ref{tab:spectral_setup},  
this velocity range covers all the emission seen in both molecules. 
The bottom left and right panels of Fig.~\ref{fig:data} show the 
integrated intensity maps for \ce{H^{13}CO+} and \ce{HNC}.


\section{Median Spectra\label{app:median_spectra}}

\begin{figure}[htb!]
    \centering
    \includegraphics[width=\columnwidth]{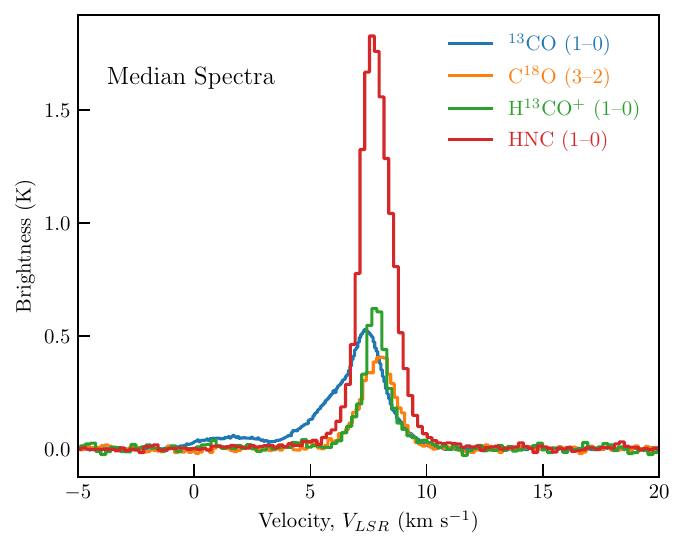}
    \caption{Typical spectra for the different tracers used calculated over the regions used in this work.}
    \label{fig:median_spectra}
\end{figure}

We show the median spectra for across the four different data cubes used in Fig.~\ref{fig:median_spectra}. 
They are determined as the median value of all the values in the cubes. 

{Given the average spectra, we estimate the optical depth of the brightest line in the sample: \ce{HNC} (1--0). 
We use the radiative transfer solution to estimate the optical depth as,
\begin{equation}
\tau(\ce{HNC}) = 
-\ln\left(1.0 - \frac{T_p (\ce{HNC})} 
                   {J_{\nu}(T_{ex}) - J_\nu(T_{cmb})} \right)~,
\end{equation}
where $T_p (\ce{HNC})$ is the observed line peak brightness, and 
$J_\nu(T)$ is the brightness temperature of a black body 
with temperature $T$ at frequency $\nu$. 
For \ce{HNC} (1--0) we assume the same excitation temperature of  $T_{ex} = 12$~K 
as used by \cite{Pineda2024-ProPStar_NGC1333_CRIR}, 
and obtain $\tau(\ce{HNC})=0.23$, which shows the emission is optically thin.
We perform the same calculations with $T_{ex} = 10$~K and obtain an optical depth of 0.31, confirming 
that the optically thin approximation is reasonable in this region.}

\section{Power-law Fit with EMCEE\label{sec:appendix_fit}}
The \verb+emcee+ fit is initialized starting from the result of the linear fit using the minimize function within the \verb+scipy+ python package \citep{2020SciPy-NMeth}.
In the  \verb+emcee+ run, we use uniform priors {$U(a, b)$, which is constant between $a$ and $b$. 
We use }
$U(1, 2.5)$ for {$A_{\ce{H^{13}CO^+}}$ and $A_{\ce{HNC}}$}, 
{$U(2.0, 3.4)$ for $\beta_{\ce{H^{13}CO^+}}$ and $\beta_{\ce{HNC}}$}, 
{$U(2, 3)$ for $f_{\rm ^{13}CO}$, 
$U(-2, -1)$ for $f_{\rm H^{13}CO^+}$, and 
$U(-1, 0.5)$ for $f_{\rm HNC}$, 
where the ranges are set to cover the best fit from the initial linear fit and without reaching 
the limits with the chains}.

We use {56} random walkers, with 50\,000 steps. We estimate an autocorrelation time of {80} steps, 
and therefore before analysing, we discard the first {600} steps and then thin the samples 
by {40} to obtain better estimates.

\begin{figure*}[b!]
    \centering
    \includegraphics[width=0.8\textwidth]{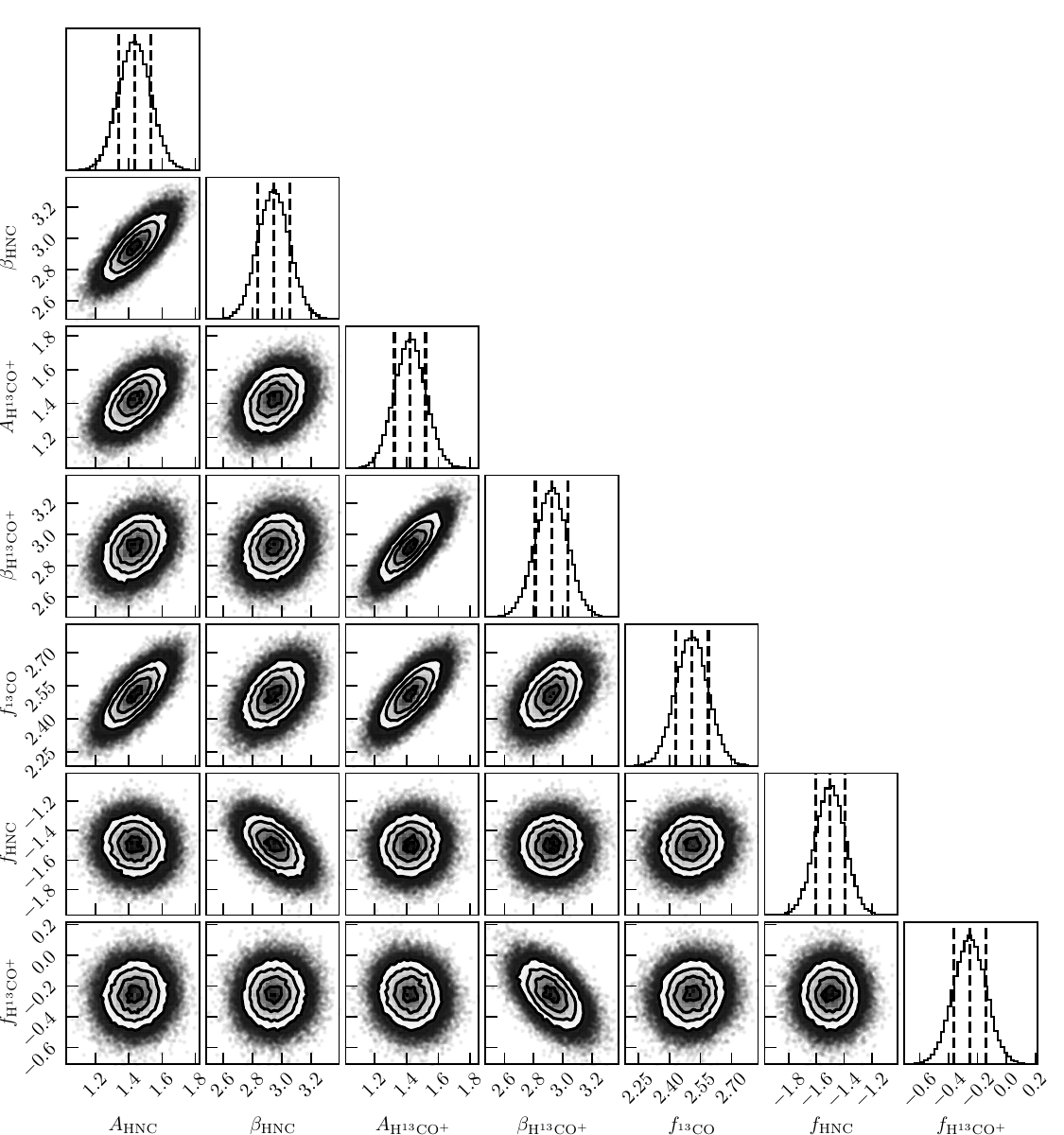}
    \caption{{Corner plot for all parameters fitted.}
    See Table~\ref{tab:power-law_fit} for a summary of the fits.}
    \label{fig:corner_plots}
\end{figure*}

The corner plots for both fits are shown in Fig.~\ref{fig:corner_plots} and summarized in Table~\ref{tab:power-law_fit}.

\end{appendix}
\end{document}